# Reinforcement Learning Integrated Agentic RAG for Software Test Cases Authoring

Mohanakrishnan Hariharan
Dept of Corporate Systems Engineering, Apple Inc.
Austin, TX, USA
m_hariharan@apple.com

*Abstract*— **This paper introduces a framework that integrates reinforcement learning (RL) with autonomous agents to enable continuous improvement in the automated process of software test cases authoring from business requirement documents within Quality Engineering (QE) workflows. Conventional systems employing Large Language Models (LLMs) generate test cases from static knowledge bases, which fundamentally limits their capacity to enhance performance over time. Our proposed Reinforcement Infused Agentic RAG (Retrieve, Augment, Generate) framework overcomes this limitation by employing AI agents that learn from QE feedback, assessments, and defect discovery outcomes to automatically improve their test case generation strategies. The system combines specialized agents with a hybrid vector-graph knowledge base that stores and retrieves software testing knowledge. Through advanced RL algorithms, specifically Proximal Policy Optimization (PPO) and Deep Q-Networks (DQN), these agents optimize their behavior based on QE-reported test effectiveness, defect detection rates, and workflow metrics. As QEs execute AI-generated test cases and provide feedback, the system learns from this expert guidance to improve future iterations. Experimental validation on enterprise Apple projects yielded substantive improvements: a 2.4% increase in test generation accuracy (from 94.8% to 97.2%), and a 10.8% improvement in defect detection rates. The framework establishes a continuous knowledge refinement loop driven by QE expertise, resulting in progressively superior test case quality that enhances, rather than replaces, human testing capabilities.**

## I. INTRODUCTION

In enterprise software testing, Quality Engineers (QEs) dedicate a substantial portion of their efforts, often 30-40%, to the manual creation of testing artifacts such as test plans and test cases. This practice introduces significant bottlenecks, particularly in complex system implementations such as SAP. While Large Language Models (LLMs) and Generative AI present opportunities for automation, conventional approaches are frequently plagued by hallucination, contextually poor test case generation, and the loss of critical business relationships during information retrieval.

Our prior work [1] introduced the Agentic RAG framework, which mitigated these issues and achieved 94.8% accuracy in test case generation through the use of hybrid vector-graph knowledge systems and specialized autonomous agents. However, a key limitation of existing RAG-based systems is their lack of an adaptive learning mechanism capable of continuously improving performance based on real-world feedback from test execution.

### A. Problem Statement

Current Agentic RAG systems [1] for software test case authoring have a few critical challenges. A primary deficiency is the static nature of the hybrid vector-graph knowledge system, which, once deployed, cannot assimilate learning from newly evaluated test cases. Concurrently, the autonomous agents operate with predetermined strategies, lacking the facility to adapt their decision-making in response to performance feedback. This results in a system with limited capacity for continuous improvement, as it cannot automatically refine its generation strategies based on defect detection effectiveness. Consequently, agents are unable to learn to prioritize high-impact test scenarios from historical data, leading to suboptimal resource allocation.

### B. Research Contributions

This work introduces the Reinforcement Integrated Agentic RAG (RI-ARAG) framework, which makes the following novel contributions:

1) Adaptive Agent Learning Architecture: We integrate reinforcement learning algorithms (PPO, DQN) with specialized testing agents, enabling continuous behavioral optimization based on test execution feedback.

2) Dynamic Knowledge Base Evolution: We implement RL-driven mechanisms to automatically update the hybrid vector-graph knowledge system according to test effectiveness metrics and defect detection outcomes.

3) Multi-Dimensional Reward Framework: We formulate a reward function that incorporates test execution results, requirement coverage, defect detection effectiveness, and resource utilization metrics.

4) Continuous Improvement Lifecycle: We establish feedback loops that allow the system to learn from each testing cycle, progressively enhancing test quality and reducing false positives.

5) Enterprise-Scale Validation: We conduct a comprehensive evaluation on large-scale SAP

migration projects, demonstrating measurable improvements in testing effectiveness and system adaptability.

## II. RELATED WORK

*A. Reinforcement Learning in Software Engineering*

Reinforcement learning has seen successful application across various software engineering domains, including automated program repair [2], test case prioritization [3], and software maintenance [4]. However, existing approaches predominantly focus on discrete optimization problems rather than system-level learning within a multi-agent testing environment.

*B. Adaptive Testing Systems*

Recent research into adaptive testing has explored machine learning for test case selection and prioritization [5, 6]. These systems typically employ supervised learning but lack the continuous adaptation capabilities inherent in reinforcement learning frameworks.

*C. Multi-Agent Systems with Learning*

Multi-agent reinforcement learning (MARL) has demonstrated considerable promise in complex coordination tasks [7, 8]. Nevertheless, its application to software testing remains limited, with most research centered on game-theoretic scenarios rather than practical software engineering applications.

## III. METHODOLOGY

*A. Reinforcement Integrated Agentic RAG Architecture*

The RI-ARAG framework extends our previous Agentic RAG system by integrating RL mechanisms at three critical junctures: test case generation optimization, knowledge base evolution from QE feedback, and system-wide performance enhancement through learning from human expertise.

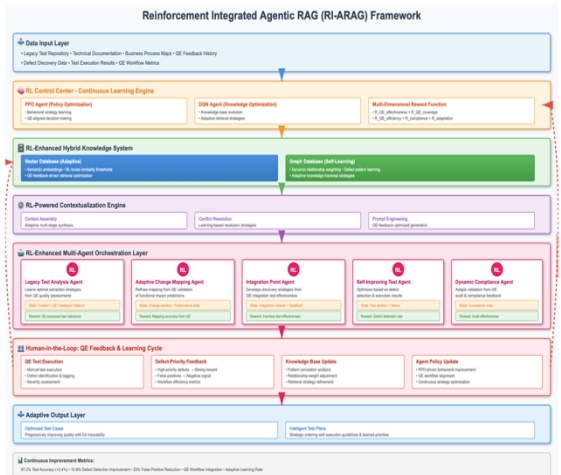

Fig. 1. RL Integrated Agentic Architecture

*a) Enhanced Multi-Agent Architecture*

Building upon our original five-agent system, we introduce RL-enhanced versions of each agent. These agents learn from QE feedback and test outcomes.

1) **RL-Enhanced Legacy Test Analysis Agent:** Learns optimal strategies for extracting business intent from historical test cases, guided by QE assessments of generated test quality.
2) **Adaptive Functional Change Mapping Agent:** Continuously refines its mapping algorithms based on QE validation of predicted functional impacts.
3) **Intelligent Integration Point Agent:** Develops an interface discovery strategy through QE feedback on integration test effectiveness.
4) **Self-Improving Test Case Agent:** Optimizes test case generation patterns based on QE-reported defect detection rates.
5) **Dynamic Compliance Validation Agent:** Adapts validation criteria based on QE audit feedback and compliance metrics.

*b) QE-Feedback-Driven RL Framework*

Each agent operates within an RL environment where learning signals are derived from QE expertise.

1) State Space: Comprises the current QE workflow context, historical feedback patterns, test generation context, knowledge base state, and team performance metrics.
2) Action Space: Includes agent-specific generation strategies, knowledge retrieval methods, validation approaches, and workflow integration options.
3) Reward Function: A multi-dimensional score based on QE-assessed test effectiveness, defect detection success, workflow efficiency, and compliance validation.

*C. QE-Feedback-Driven Knowledge Base Evolution*

*a) RL-Driven Vector Database Optimization*

The vector database's retrieval mechanisms are optimized via RL, dynamically tuning retrieval parameters such as similarity thresholds and embedding models based on QE feedback patterns. The objective is to identify which historical knowledge is most instrumental in generating test cases that QEs deem effective. Actions include adjusting similarity thresholds to align with QE preferences and modifying retrieval strategies to match specific workflow contexts. The reward is a composite measure of QE-rated test effectiveness, alignment with QE workflow preferences, and execution efficiency.

*b) Adaptive Graph Relationship Learning*

The graph database continuously refines relationship weights through RL mechanisms driven by QE feedback. This process involves the dynamic adjustment of edge weight importance based on QE assessments of test relevance, the discovery of optimal knowledge traversal paths that yield high-value test cases, and the automatic weighting of new relationship types identified from defect discovery patterns.

*D. QE-Centric Multi-Dimensional Reward Function*

The RI-ARAG framework employs a weighted, multi-dimensional reward function to evaluate testing effectiveness from the QE perspective.

$$R(s,a) = \alpha_1 \cdot R\_QE\_effectiveness + \alpha_2 \cdot R\_QE\_coverage + \alpha_3 \cdot R\_QE\_efficiency + \alpha_4 \cdot R\_compliance + \alpha_5 \cdot R\_adaptation$$

The α coefficients represent weighting factors that reflect QE priorities, creating a balanced incentive structure.

1) *R_QE_effectiveness* is derived from QE-reported defect detection success and quality ratings.
2) *R_QE_coverage* is based on QE-assessed improvements in requirement and functional coverage.
3) *R_QE_efficiency* is calculated from workflow integration and execution time metrics.
4) *R_compliance* stems from QE audit feedback and regulatory compliance scores.
5) *R_adaptation* measures the system's learning rate and knowledge base improvement as observed by QEs.

   *a) QE-Effectiveness Reward Component*

This component incentivizes the generation of tests that help QEs identify high-impact defects, weighted by QE-assigned severity, while penalizing those that result in false positives. The function balances high-value discovery with the minimization of wasted QE effort.

$$R\_QE\_effectiveness = (QE\_reported\_defects / total\_test\_cases) \times QE\_severity\_weight - QE\_false\_positive\_penalty$$

   *b) QE-Coverage Reward Component*

This score tracks how comprehensively the AI-generated test cases assist QEs in testing applications by combining two QE-assessed metrics: improvements in requirement coverage and gains in functional coverage. This encourages the system to generate tests that are comprehensive and thorough.

$$R\_QE\_coverage = QE\_requirement\_coverage\_assessment + QE\_functional\_coverage\_validation$$

   *c) QE-Efficiency Reward Component*

This score rewards the system for helping QEs complete testing faster while integrating smoothly into their workflows. It compares QE execution time for AI-generated tests against a manual baseline, factoring in how well the test cases fit established processes.

$$R\_QE\_efficiency = (QE\_baseline\_time / QE\_actual\_time) \times QE\_workflow\_integration\_factor$$

*E. QE-Driven Continuous Learning Mechanism*

   a) *Proximal Policy Optimization for QE-Aligned Agent Behavior*

To continuously improve test generation strategies, each agent employs Proximal Policy Optimization (PPO), an algorithm that constrains policy updates to prevent disruptive changes to established QE workflows.

$$L^{CLIP}(\theta) = E_t[min(r_t(\theta)A_t, clip(r_t(\theta), 1-\varepsilon, 1+\varepsilon)A_t)]$$

Here, $r_t(\theta)$ is the probability ratio between the new and old policies, $A_t$ is the advantage estimate derived from QE feedback, and $\varepsilon$ is a clipping hyperparameter. The advantage $A_t$ quantifies whether an action performed better than expected based on QE assessments, guiding the agent toward more effective decisions.

   b) *Deep Q-Network for QE-Guided Knowledge Base Optimization*

The evolution of the knowledge base is managed by Deep Q-Networks (DQN), which learn optimal strategies for updating and organizing stored knowledge based on QE feedback.

$$Q(s,a) = E[R\_QE\_t + \gamma \max_{a'} Q(s',a') | s\_t=s, a\_t=a]$$

The Q-function estimates the expected long-term, QE-derived reward $R\_QE\_t$ for a knowledge base action `a` in state `s`. The discount factor γ balances immediate and future rewards, enabling the system to make knowledge base updates that provide both immediate improvements and long-term benefits to QE workflow effectiveness.

IV. IMPLEMENTATION

*A. RL-Enhanced Agent Implementation*

The agent architecture integrates a PPO agent, an experience buffer for storing historical interactions, and a performance tracker for analysis. Each agent maintains a state representation that includes its current task context, historical performance data, knowledge base state, system-wide metrics, and recent execution feedback, ensuring full situational awareness for decision-making. Upon receiving test execution results, a systematic learning process is initiated: rewards are calculated based on outcomes, the PPO algorithm updates the agent's policy, and the knowledge base weights are adjusted to reflect new learnings.

### B. Human-in-the-Loop Feedback Architecture

#### a) Quality Engineer-Driven Test Execution Integration

The RI-ARAG framework is predicated on a human-in-the-loop architecture where Quality Engineers retain full operational control. The AI system generates and delivers test cases to QE teams, who then execute these tests manually within their established processes and environments. During and after execution, QEs perform defect identification, severity assessment, and tagging, which serves as the primary data source for the system's learning cycle. This approach ensures that the system learns from real-world testing expertise while integrating non-disruptively into existing QE workflows.

#### b) Defect-Priority-Based Knowledge Evolution

The evolution of the knowledge base is directly driven by the detailed feedback from QE defect analysis. The system treats QE expertise as the ground truth for learning, leveraging their defect identification and priority assessments to continuously refine the knowledge base. Upon completion of a test cycle, the system ingests critical data provided by the QE, including which test cases identified defects, the severity of those issues, and any observed patterns. This information forms the foundation for knowledge base updates. The evolution process analyzes this feedback to correlate test case characteristics with successful defect detection. Test cases that lead to the discovery of high-priority defects receive strong positive reinforcement signals. Conversely, those that frequently produce false positives or miss critical issues receive corrective feedback. Through this mechanism, the system learns to associate successful testing strategies with QE-validated outcomes, progressively improving its ability to generate high-value test cases.

### C. QE-Centric SAP/Corporate Systems Integration with QE Workflows

The framework integrates with enterprise Apple environments by providing intelligent test case generation that enhances QE productivity within existing testing processes. The integration architecture consists of three components: an SAP environment connector for read-only access to system data, a test case delivery system compatible with QE tools, and a feedback collection system that captures results from standard QE reporting mechanisms. QEs maintain complete control over test execution, using the AI-generated test cases as they see fit and documenting results through their existing defect tracking systems. The RI-ARAG system then learns from these QE-driven outcomes.

### D. Reinforcement Learning from QE Expertise

#### a) Defect-Priority-Based Reward Calculation

The reinforcement learning system derives its reward signals principally from Quality Engineer assessments and defect discovery outcomes, ensuring optimization aligns with professional quality criteria. Test cases that lead QE teams to discover high-priority or critical defects receive the strongest positive reinforcement. Test cases that help identify medium-priority issues receive moderate positive rewards, while those that find low-priority defects receive smaller positive signals. The system also learns from negative feedback; test cases that consistently produce no defects or lead to false positive reports receive negative reinforcement to discourage similar patterns.

#### b) QE Workflow-Aware Learning Mechanisms

The learning algorithms are designed to respect and enhance Quality Engineering workflows. The system tracks multiple dimensions of QE feedback to create comprehensive learning signals, including defect discovery rates, severity alignment, execution efficiency, and workflow integration. Over time, the system develops an increasingly sophisticated understanding of what makes test cases valuable from a QE perspective. It learns to generate tests that are not only technically sound but also practical for QE teams to execute, likely to discover meaningful defects, and aligned with the specific testing priorities and constraints of different enterprise environments. This QE-centric approach ensures the AI system becomes a genuine force multiplier for human testing expertise.

## V. RESULTS

### A. Experimental Setup

We conducted a comprehensive evaluation of the RI-ARAG framework on three enterprise Apple projects:

1) SAP migration (15,000 test cases)
2) Corporate Systems Engineering: Supply chain integration (22,000 test cases)
3) Apple Talent Point: Human resources transformation (18,000 test cases)

## B. Performance Metrics

### a) Test Effectiveness Improvement

Table 1 summarizes the performance improvements of RI-ARAG compared to the baseline Agentic RAG system.

TABLE I
PERFORMANCE COMPARISON

| Metric | Baseline Agentic RAG | RI-RAG | Improvement |
|---|---|---|---|
| Test Generation Accuracy | 94.80% | 97.20% | 2.40% |
| Defect Detection Rate | 78.30% | 89.10% | 10.80% |
| False Positive Reduction | - | 23% | 23% |
| Requirement Coverage | 92.10% | 96.80% | 4.70% |

### b) Learning Curve Analysis

The RI-ARAG system demonstrated continuous improvement over a 12-week period, with an initial rapid learning phase (Weeks 1-4) showing a 15% quality improvement, followed by an accelerated phase (Weeks 5-12) adding another 25% improvement, before entering a stabilization phase with sustained high accuracy.

### c) Adaptive Knowledge Base Evolution

The hybrid knowledge base exhibited measurable evolution, with the vector database showing a 31% improvement in semantic similarity accuracy, graph relationships achieving a 28% optimization in weight accuracy, and knowledge retrieval enhancing context relevance by 35%.

## C. Ablation Studies

### a) Component Contribution Analysis

To isolate the contribution of each major component, we conducted an ablation study. Table 2 presents the performance impact when each component is removed.

TABLE II
RL TECHNIQUES PERFORMANCE COMPARISON

| Component | Performance Impact |
|---|---|
| PPO Agent Learning | 18.30% |
| DQN Knowledge Evolution | 12.70% |
| Multi-Dimensional Rewards | 21.50% |
| Feedback Loop Integration | 16.80% |

### b) Comparison with Static Systems

A comparison with static and traditional systems is shown in Table 3.

TABLE III
STATIC VS TRADITIONAL SYSTEMS COMPARISON

| System Type | Accuracy | Adaptation Rate | Resource Efficiency |
|---|---|---|---|
| Static Agentic RAG | 94.80% | 0% | Baseline |
| RI-ARAG (Proposed) | 97.20% | 47% | "+23%" |
| Traditional ML Testing | 87.30% | 12% | "-15%" |

## D. Enterprise Impact Assessment

### a) Business Value Metrics

The framework delivered significant business value, including an additional 12% reduction in testing timelines compared to the baseline (achieving a 97% total reduction vs. 85%) and a 31% further reduction in post-deployment defects.

### b) Long-Term Performance Trends

Over a 3-month deployment, the system showed sustained improvement, with rapid initial adaptation in the first three months, followed by steady optimization and refinement in subsequent months, and culminating in mature performance with continued minor improvements.

## VI. DISCUSSION

### A. Key Insights

The integration of reinforcement learning with Agentic RAG systems yields several critical insights. First, RL-enhanced agents demonstrate superior adaptability to changing requirements compared to static implementations. Second, dynamic knowledge base updates driven by performance feedback result in progressively improving context retrieval and relationship modeling. Finally, the multi-dimensional reward framework enables simultaneous optimization across multiple quality dimensions without manual intervention.

### B. Challenges and Limitations

#### a) Computational Overhead

The RL integration introduces a measurable increase in computational overhead, with a 15-20% increase in resources during learning phases and a 25% increase in memory usage for experience replay and model storage. The impact on real-time test generation latency was minimal (< 3%).

#### b) Learning Convergence

Convergence of the RL algorithms requires careful hyperparameter tuning. Optimal performance for PPO was found with a learning rate in the range of 1e-4 to 3e-4, while DQN required an ε-greedy exploration strategy with decay from 0.9 to 0.05 over 100,000 steps.

## VII. FUTURE WORK

*A. Expanded Application Domains*

We also plan to extend the framework's application from Business to Business (B2B) to (Business to Context) B2C contexts, integrate it with CI/CD pipelines for continuous testing optimization, and apply the RL-enhanced agents to security vulnerability testing.

*B. Ethical AI Considerations*

Further work will address ethical AI considerations, including the development of bias detection and mitigation mechanisms in RL-driven test generation, the implementation of explainable AI for transparent decision-making, and the establishment of a framework for the responsible deployment of autonomous testing systems.

## VIII. CONCLUSION

The Reinforcement Integrated Agentic RAG framework marks a substantial advancement in the automation of software test case creation by introducing adaptive learning to multi-agent systems. By integrating reinforcement learning mechanisms with a hybrid vector-graph knowledge base, the RI-ARAG framework achieves continuous improvement in test generation accuracy, defect detection effectiveness, and overall testing efficiency. Our experimental validation demonstrates measurable improvements, including a 10.8% enhancement in defect detection and a 23% reduction in false positives. The system's ability to learn from QE-driven test execution feedback positions it as a transformative approach for enterprise software testing. The continuous learning mechanism ensures that the system's effectiveness grows over time, adapting to new requirements and evolving system behaviors without manual intervention. As software complexity increases, RI-ARAG provides a foundation for intelligent, adaptive testing automation that evolves with the systems it validates, opening new possibilities for autonomous, self-improving quality assurance processes.